\documentclass[conference]{IEEEtran}
\IEEEoverridecommandlockouts
\newtheorem{theorem}{Theorem}

\newtheorem{lemma}[theorem]{Lemma}
\newtheorem{remark}[theorem]{Remark}

\newtheorem{definition}[theorem]{Definition}

\usepackage{cite}

\ifCLASSINFOpdf
  \usepackage[pdftex]{graphicx}
\else
\fi
\usepackage{color}

\usepackage[cmex10]{amsmath}
\usepackage{amssymb}

\usepackage{algorithm}
\usepackage{algpseudocode}

\usepackage[caption=false,font=footnotesize]{subfig}

\usepackage{url}

\usepackage{array}
\begin{document}
\title{Design of Improved Quasi-Cyclic Protograph-Based Raptor-Like LDPC Codes for Short Block-Lengths\thanks{Research is supported in part by National Science Foundation (NSF) grant CCF-1618272. Any opinions, findings, and conclusions or recommendations expressed in this material are those of the author(s) and do not necessarily reflect the views of the NSF. Research was carried out in part at the Jet Propulsion Laboratory (JPL), California Institute of Technology, under a contract with NASA. This work used computational and storage services associated with the Hoffman2 Shared Cluster provided by UCLA Institute for Digital Research and Education's Research Technology Group.}}

\author{\IEEEauthorblockN{Sudarsan V. S. Ranganathan\IEEEauthorrefmark{1}, 
Dariush Divsalar\IEEEauthorrefmark{2}, and
Richard D. Wesel\IEEEauthorrefmark{1}}
\IEEEauthorblockA{\IEEEauthorrefmark{1}Department of Electrical Engineering, 
University of California, Los Angeles, Los Angeles, California 90095\\
\IEEEauthorrefmark{2}Jet Propulsion Laboratory, California Institute of Technology, Pasadena, California 91109\\Email: sudarsanvsr@ucla.edu, Dariush.Divsalar@jpl.nasa.gov, wesel@ucla.edu}
}



\maketitle
\begin{abstract}
Protograph-based Raptor-like low-density parity-check codes (PBRL codes) are a recently proposed family of easily encodable and decodable rate-compatible LDPC (RC-LDPC) codes. These codes have an excellent iterative decoding threshold and performance across all design rates. PBRL codes designed thus far, for both long and short block-lengths, have been based on optimizing the iterative decoding threshold of the protograph of the RC code family at various design rates. 

In this work, we propose a design method to obtain better quasi-cyclic (QC) RC-LDPC codes with PBRL structure for short block-lengths (of a few hundred bits). We achieve this by maximizing an upper bound on the minimum distance of any QC-LDPC code that can be obtained from the protograph of a PBRL ensemble. The obtained codes outperform the original PBRL codes at short block-lengths by significantly improving the error floor behavior at all design rates. Furthermore, we identify a reduction in complexity of the design procedure, facilitated by the general structure of a PBRL ensemble. 
\end{abstract}
%
\IEEEpeerreviewmaketitle
\section{Introduction and Background}
\label{sec_intro}
Protograph-based low-density parity-check (LDPC) codes \cite{Proto}, \cite{5174517} are a class of codes amenable to tractable analysis and design procedures. Protograph quasi-cyclic LDPC (QC-LDPC) codes \cite{1317123}, a class of protograph codes, have parity-check matrices composed of circulant permutation matrices (CPMs) and permit very low complexity decoder implementations \cite{5174517}.

The presence of CPMs in protograph QC-LDPC codes enables us to understand how the connections in the protograph affect the girth and minimum distance of such codes. Fossorier \cite{1317123}, Karimi and Banihashemi \cite{6475181}, and others analyze the girth of a protograph QC-LDPC code by examining the protograph of the code. More pertinent to this paper are \cite{6145509} and \cite{6482231}. Smarandache and Vontobel \cite{6145509} derive an upper bound on the minimum distance of any QC-LDPC code that can be obtained from a protograph. Butler and Siegel \cite{6482231} extend the results of \cite{6145509} to QC-LDPC codes based on punctured protographs.

Protograph-based Raptor-like LDPC codes (PBRL codes) are a class of easily encodable rate-compatible (RC) LDPC code families proposed by Chen, Vakilinia et al.\ in \cite{7045568}. PBRL code families have an excellent iterative decoding threshold \cite{910578}, \cite{5174517} and performance across all rates for which they are designed. In \cite{7045568}, the authors design PBRL protographs for long and short block-lengths by optimizing the iterative decoding threshold of the protograph at each rate. They show that PBRL QC-LDPC code families can outperform other RC-LDPC codes in the literature, both at short ($\approx 1000$ information bits) and long ($\approx 16000$ information bits) block-lengths.

This paper considers the design of RC-LDPC codes for very short block-lengths ($\approx 200$ information bits). While the iterative decoding threshold is the correct design metric to use for code design at long block-lengths, minimum distance is more important at short block-lengths. 

One contribution of this paper is a new PBRL design approach. Given a set of design rates, we design protographs for PBRL ensembles by maximizing, at each rate, the upper bounds on the minimum distance that were derived in \cite{6145509} and \cite{6482231}. The resulting PBRL QC-LDPC code families outperform the ones designed by optimizing the iterative decoding threshold at each rate. The complexity of computing the aforementioned upper bounds increases quickly with the size of the protographs. A second contribution of this paper is to leverage the structure of PBRL protographs to identify a significant reduction in the complexity of the design procedure. 

The paper is organized as follows: Section \ref{sec_design} provides the design procedure, derives the reduction that is possible in the computational complexity of the design procedure, and discusses design examples. Section \ref{sec_results} shows simulation results. Section \ref{sec_conclusion} concludes the paper. 
\section{Designing PBRL Ensembles by Maximizing an Upper Bound on the Minimum Distance}
\label{sec_design}
A PBRL ensemble is defined by its protomatrix $P$ that has the following general form:
\begin{align}
\label{eq_pbrl_general}
P = \begin{bmatrix}
P_{\mathsf{HRC}} & 0 \\
P_{\mathsf{IRC}} & I
\end{bmatrix}_{n_c \times n_v}
\end{align}
Here, $0$ and $I$ refer to the all-zeros and identity matrices of appropriate size. The highest-rate code $\left(\mathsf{HRC}\right)$ of the rate-compatible protomatrix is represented by $P_{\mathsf{HRC}}$, which is of size $n_{c_H} \times n_{v_H}$. The variable nodes of the protomatrix containing the identity matrix in \eqref{eq_pbrl_general} represent the incremental redundancy symbols of $P$. Some variable nodes of a protomatrix could also be punctured. The design rate is $R \triangleq \left(n_v-n_c\right)/n_t$ for a protomatrix with $n_c$ check nodes, $n_v$ variable nodes, and $n_t$ of the $n_v$ variable nodes that are transmitted.

A protomatrix is lifted (see \cite{5174517}, \cite{7045568}) to obtain an LDPC code of block-length that is a multiple of $n_t$. The use of circulant permutation matrices while lifting yields QC-LDPC codes, which are practical and are the subject of this paper. 
\subsection{Design Method}
\label{subsec_design_method}
The design of a PBRL protomatrix consists of two steps: First, we choose the $\mathsf{HRC}$ part, $P_{\mathsf{HRC}}$, as a protomatrix by itself. Then, we obtain the $\mathsf{IRC}$ part, $P_{\mathsf{IRC}}$, one row at a time. In the original work on PBRL codes by Chen, Vakilinia et al.\ \cite{7045568}, the authors first choose an $\mathsf{HRC}$ part with a degree distribution and an acceptable iterative decoding threshold. Then they design each row of $P_{\mathsf{IRC}}$ successively to optimize the iterative decoding threshold of the protomatrix up to that rate while designing the row, keeping all previously obtained rows fixed. The best known families of RC-LDPC codes at both short and long block-lengths are the PBRL codes as designed with the heuristics proposed by Chen, Vakilinia et al.\ in \cite{7045568}. 

At short block-lengths, minimum distance can be more important than the iterative decoding threshold as a criterion to use while designing LDPC codes. A key feature of QC-LDPC codes based on protomatrices is that the minimum distance of any such code obtained from a protomatrix is upper bounded by a constant that depends only on the protomatrix. In order to state the upper bounds, which were derived in earlier works, we need the definition of the permanent of a square matrix.

\begin{definition}[Permanent]
\label{def_permanent}
Denote the set $\{1, 2, \dots, \ell\}$ by $[\ell]$. The permanent of an $\ell \times \ell$ square matrix $A$ with elements $a_{i,j}, i \in [\ell], j \in [\ell]$ over some commutative ring is defined as
\begin{align}
\label{eq_permanent_definition}
\mathsf{perm} (A) = \sum_\sigma \prod_{1 \le j \le \ell} a_{j, \sigma(j)} = \sum_\sigma \prod_{1 \le j \le \ell} a_{\sigma(j), j},
\end{align}
where $\sigma$ refers to a permutation and the summation is over all permutations of $[\ell]$. The permanent, although it looks deceptively similar to the determinant, is harder to compute than the determinant \cite{6352911}. While the arithmetic complexity of computing the determinant is $O\left(\ell^3\right)$, the most efficient algorithm known to compute the permanent of any square matrix, due to Ryser \cite{CM_Ryser}, is of complexity $\Theta\left(\ell \cdot 2^\ell\right)$. 
\end{definition}

\begin{theorem}[Upper bound for unpunctured protomatrices; Theorem 8 of \cite{6145509}]
\label{theorem_theorem_unpunctured}
Let a protomatrix $P$ with a positive design rate and no punctured variable nodes be of size $n_c \times n_v$. If $S \subseteq \left[n_v\right]$, denote by $P_S$ the sub-matrix of $P$ formed by the columns indexed by elements of $S$. Then, any QC-LDPC code $\mathcal{C}$ obtained from the protomatrix $P$ has a minimum distance $d_{\text{min}}(\mathcal{C})$ that is upper bounded as
\begin{align}
\label{eq_first_theorem}
d_{\text{min}}(\mathcal{C}) \le \underset{S \subseteq [n_v], |S| = n_c + 1}{\text{min$^\ast$}} \sum_{i \in S} \mathsf{perm}\left(P_{S \setminus i}\right),
\end{align}
where $|\cdot|$ refers to the cardinality of a set, $S \setminus i$ is shorthand for $S \setminus \{i\}$, and min$^\ast$ returns the smallest non-zero value in a set of non-negative values with at least one positive value or $+\infty$ if the set is $\{0\}$. Note that permanents computed from sub-matrices of a protomatrix are always non-negative.
\end{theorem}

\begin{theorem}[Upper bound for punctured protomatrices; Theorem 9 of \cite{6482231}]
\label{theorem_theorem_punctured}
Let a punctured protomatrix $P$ with a positive design rate less than 1 be of size $n_c \times n_v$. Let the set of punctured variable nodes, a subset of $[n_v]$, be denoted $\mathcal{P}$. Denote any punctured QC-LDPC code that can be obtained from $P$ by $\mathcal{C}'$ and the unpunctured version of the code $\mathcal{C}'$ by $\mathcal{C}$. Then, provided that $\mathcal{C}$ and $\mathcal{C}'$ have the same number of codewords in their codebooks (dimensionality), $\mathcal{C}'$ has a minimum distance $d_{\text{min}}(\mathcal{C}')$ that is upper bounded as
\begin{align}
\label{eq_second_theorem}
d_{\text{min}}(\mathcal{C}') \le \underset{S \subseteq [n_v], |S| = n_c + 1}{\text{min$^\ast$}} \sum_{i \in S \setminus \mathcal{P}} \mathsf{perm}\left(P_{S \setminus i}\right).
\end{align}
\end{theorem}

With Theorems \ref{theorem_theorem_unpunctured} and \ref{theorem_theorem_punctured} in hand, we propose the following PBRL ensemble search procedure:
\begin{enumerate}
\item Choose an $\mathsf{HRC}$ matrix of size $n_{c_H} \times n_{v_H}$ with a desired degree distribution and complexity constraint. A common complexity constraint is to limit the weight of each column in the protomatrix.
\item $\mathsf{IRC}$ design: Select the next row of the protomatrix from a set of candidate rows to maximize the upper bound on the minimum distance via Theorem \ref{theorem_theorem_punctured} or \ref{theorem_theorem_unpunctured} (depending upon whether there are punctured nodes or not). If there are multiple candidates with the best upper bound, then select one at random.
\item Go to Step 2) if another row of $\mathsf{IRC}$ is required. Otherwise, terminate the search procedure. 
\end{enumerate}

It is not known, in general, whether the upper bounds of \eqref{eq_first_theorem} or \eqref{eq_second_theorem} are achievable. But our design procedure yields better RC QC-LDPC code families at short block-lengths than the design based on optimizing the iterative decoding thresholds. 

For a punctured protomatrix, care must be taken to ensure that not too many variable nodes are punctured. Otherwise, the dimensionality requirement in Theorem \ref{theorem_theorem_punctured} may be violated. 
\subsection{Lowering the Complexity of the Design Procedure}
\label{subsec_complexity}
In this subsection, we leverage the general structure of the protomatrix of a PBRL ensemble in \eqref{eq_pbrl_general} to reduce the complexity of computing the upper bounds in \eqref{eq_first_theorem} or \eqref{eq_second_theorem}. 

Assume that we have a PBRL protomatrix of size $n_c \times n_v$ with an $\mathsf{HRC}$ part that is of size $n_{c_H} \times n_{v_H}$. Assume also that the protomatrix has no punctured variable nodes\footnote{We consider the case when the protomatrix has punctured variable nodes in the Appendix.}. With these assumptions, computation of the upper bound for the protomatrix as given in \eqref{eq_first_theorem} requires computing $\binom{n_v}{n_c + 1}\cdot \left(n_c + 1\right)$ permanents, each of size $n_c \times n_c$. The complexity of Ryser's algorithm to compute the required permanents increases quickly while constructing the $\mathsf{IRC}$ part of a PBRL ensemble. Our following result leads to a significant reduction in both the number of permanents that need to be computed and the size of each permanent to be computed.
\begin{theorem}
\label{theorem_main_result}
Let a PBRL protomatrix $P$ of size $n_c \times n_v$ with no punctured variable nodes have a positive design rate, i.e.\ $n_v > n_c$. Let the $\mathsf{HRC}$ part be of size $n_{c_H} \times n_{v_H}$. Assume that the upper bound in $\eqref{eq_first_theorem}$ for $P$ is a positive integer. Then, the same upper bound can be obtained with at most $\binom{n_{v_H}}{n_{c_H} + 1}\cdot \left(n_{c} + 1\right)$ permanents, each of size at most $\left(n_{c_H} + 1\right) \times \left(n_{c_H} + 1\right)$. 
\end{theorem}

Before we provide the proof, we comment on the reduction in complexity of computing \eqref{eq_first_theorem}. The complexity of computing each permanent would now depend only on the number of check nodes in the $\mathsf{HRC}$ part, $n_{c_H}$. Also, the dominating factor in the expression for number of permanents to be computed is the binomial coefficient, which again would now depend only on the size of the $\mathsf{HRC}$ part, $n_{c_H} \times n_{v_H}$, and not on the size of the entire protomatrix. 

\begin{IEEEproof}
Let us first consider the case when $S \subseteq \left[n_v\right]$, $|S| = n_c + 1$ contains the last $n_v - n_{v_H}$ columns, i.e.\ the columns that comprise the incremental redundancy variable nodes of the protomatrix and have an identity matrix of size $\left(n_{v} - n_{v_H}\right) \times \left(n_{v} - n_{v_H}\right)$. Note that $n_v - n_{v_H} = n_c - n_{c_H}$. The $n_c + 1$ chosen columns form a sub-matrix with structure that can be written as:
\begin{align}
\label{eq_structure_11}
P_S = \left[c_1~c_2~\cdots~c_{n_{c_H}+1}~|~P_{\mathsf{IR}}\right], 
\end{align}
where $c_i$ are the columns chosen from the initial $n_{v_H}$ columns of $P$ and $P_{\mathsf{IR}}$ has the following structure:
\begin{align}
\label{eq_IR_structure}
P_{\mathsf{IR}} = \begin{bmatrix}
0\\
I
\end{bmatrix}_{\left\{n_c \times \left(n_v - n_{v_H}\right)\right\}}
\end{align}
Because each column in $P_{\mathsf{IR}}$ contains only a single 1, the complexity of computing each of the $n_c + 1$ required permanents is at most the size of computing the permanent of an $\left(n_{c_H}+1\right) \times \left(n_{c_H}+1\right)$ sub-matrix (when the removed column is from $P_{\mathsf{IR}}$). When the removed column is not from $P_{\mathsf{IR}}$, the complexity is the size of computing the permanent of an $n_{c_H} \times n_{c_H}$ sub-matrix since the product is zero for permutations that select elements not in the $\mathsf{HRC}$ rows of $c_i, 1 \le i \le n_{c_H}+1$. Furthermore, there are $\binom{n_{v_H}}{n_{c_H} + 1}$ sets $S \subseteq \left[n_v\right]$ of size $n_c + 1$ that contain $P_{\mathsf{IR}}$.

Now let us consider the general set $S$ of $n_c+1$ columns in $\left[n_v\right]$. First, let us assume that
\begin{align}
\label{eq_assume_sum}
\sum_{i \in S} \mathsf{perm}\left(P_{S \setminus i}\right) > 0,
\end{align}
which implies that at least one of the $n_c + 1$ permanents is positive. Denote by $P'$ one such $n_c \times n_c$ sub-matrix of $P_S$ with a positive permanent. There exists a permutation denoted $\sigma^*$ that has a positive product in \eqref{eq_permanent_definition} when computed for the matrix $P'$. Assume the following definition of a permanent:
\begin{align}
\label{eq_second_definition_perm}
\mathsf{perm}\left(P'\right) = \sum_\sigma \prod_{1 \le j \le n_c} p'_{\sigma(j), j},
\end{align}
where $p'_{i,j}$ denotes the entries of $P'$. Consider all columns indexed by $j \in \left[n_c\right]$ such that $\sigma^*(j) > n_{c_H}$. There are $n_c - n_{c_H} = n_v - n_{v_H}$ such columns. Replace all these columns by the columns of the sub-matrix $P_\mathsf{IR}$ (whenever possible), in the following manner: Replace column $j$ whose $\sigma^*(j) = j' > n_{c_H}$ with the column in $P_\mathsf{IR}$ whose only non-zero element, 1, is present in row $j'$, unless the column from $P_{\mathsf{IR}}$ is already in the set $S$ of $n_c + 1$ columns under consideration. Call the newly obtained matrix $P'_1$. The sub-matrix $P'_1$ has a permanent that is positive and is at most the value of the permanent of $P'$ due to the following reasons: Permutation $\sigma^*$ yields a positive product with $P'_1$ because the replacements (whenever possible) only lead to non-zero entries at locations $\left(\sigma^*(j),j\right): \sigma^*(j) > n_{c_H}$. Furthermore, each permutation $\sigma$ that yields a positive product $\prod_{i \in \sigma} p'_{\sigma(j),j}$ in $P'$ (including $\sigma^*$) yields a product with $P'_1$ that is upper bounded by the product computed with $P'$. 

Let us denote the matrix $\left[c_1 ~c_2~ c_3~ \cdots ~c_{n_{c_H}+1} ~|~ P_\mathsf{IR}\right]$ by $P''$, where $c_1, c_2, \dots, c_{n_{c_H}+1}$ are the columns in $P_S$ chosen from the first $n_{v_H}$ columns of the protomatrix $P$ and were either in $P'$ and not replaced to obtain $P'_1$ or was not in $P'$. Let us denote by $S''$ the columns of $P$ that lead to $P''$. It is now straight-forward to see from the composition of the matrices $P_S$ and $P_{S''}$ that 
\begin{align}
\label{eq_main_completion}
\sum_{i \in S} \mathsf{perm}\left(P_{S \setminus i}\right) \ge \sum_{i \in S''} \mathsf{perm}\left(P_{S'' \setminus i}\right) > 0.
\end{align}

We now consider the final case of the general set of $n_c + 1$ columns whose $n_c + 1$ permanents sum to zero. Recall that the statement of the theorem assumes that $P$ has a positive upper bound in \eqref{eq_first_theorem}. Therefore, we may ignore such a case unless those columns contain $P_{\mathsf{IR}}$, in which case we would compute the sum of $n_c + 1$ permanents according to the complexity as shown already.

This completes the proof as the above shows it suffices to consider subsets $S$ that always contain the columns of $P_{\mathsf{IR}}$.
\end{IEEEproof}

The requirement that the protomatrix have a finite upper bound on its minimum distance (as assumed in Theorem \ref{theorem_main_result} and its counterpart in the Appendix for a protomatrix with punctured variable nodes) can be satisfied while designing every row of the $\mathsf{IRC}$ part via the following observation:
\begin{lemma}
\label{lemma_auxiliary_lemma}
Let a PBRL protomatrix $P$ of size $n_c \times n_v$ with a positive design rate have an $\mathsf{HRC}$ part of size $n_{c_H} \times n_{v_H}$. Let the $\mathsf{HRC}$ part, as a protomatrix by itself, have a positive and finite upper bound $d_\mathsf{HRC}$ as computed using \eqref{eq_first_theorem} or \eqref{eq_second_theorem}. Then the upper bounds for each new row $i=n_{c_H}+1,n_{c_H}+2,\dots, n_{c}$ added to obtain $P$, irrespective of the chosen candidates for the rows, are non-decreasing and are lower bounded by $d_\mathsf{HRC}$.
\end{lemma}
\begin{IEEEproof}
Let us consider the design of the first row of the $\mathsf{IRC}$ part and assume that the protomatrix has no punctured columns. Assume that there is no non-zero integer in the new $\left(n_{c_H}+1\right)^{\text{th}}$ row except the required $1$ from the $P_\mathsf{IR}$ part at entry $\left(n_{c_H}+1, n_{v_H} +1 \right)$. For any set of $n_{c_H} +2$ columns $S$ that does not include the $\left(n_{v_H}+1\right)^{\text{th}}$ column, the sum of the $n_{c_H} +2$ permanents is zero. For any other set that includes the $\left(n_{v_H}+1\right)^{\text{th}}$ column, the sum of the permanents is equal to one of the sums of $n_{c_H}+1$ permanents computed to find the upper bound for the $\mathsf{HRC}$ part. Now, if the new row is designed to have non-zero entries in the columns not in $P_{\mathsf{IR}}$, the upper bound can only increase or remain the same. 

Similar arguments follow if the protomatrix has punctured columns. This completes the proof as the above arguments can then be successively applied to each new row.
\end{IEEEproof}


\subsection{Design Examples}
\label{subsec_design_examples}
In this subsection we design PBRL protomatrices according to the new design method we have proposed. We assume the following $\mathsf{HRC}$ matrix for all our designs:
\begin{align}
\label{eq_chosen_HRC}
\begin{bmatrix}
2&1&2&1&2&1&2&1\\
1&2&1&2&1&2&1&2
\end{bmatrix}
\end{align}
We consider both a punctured and an unpunctured version of the matrix in \eqref{eq_chosen_HRC} in our examples. The punctured version has the first variable node punctured. Hence the design rate we start with is either $6/8$ (unpunctured) or $6/7$ (punctured). 

Because of Lemma \ref{lemma_auxiliary_lemma}, choosing an $\mathsf{HRC}$ with a non-zero and finite upper bound \eqref{eq_first_theorem} or \eqref{eq_second_theorem} is sufficient to then use our results on reducing the complexity of search procedure to design the $n_c \times n_v$ protomatrix $P$. The $\mathsf{HRC}$ matrix in \eqref{eq_chosen_HRC} has an upper bound of 12 when none of its variable nodes are punctured and an upper bound of 8 when its first column is punctured. 

\begin{remark}[Design constraints]
\label{remark_design_constraints}
We constrain the last $n_c - n_{c_H}$ rows of the protomatrix to have a weight of exactly 4 and do not allow any non-zero integer other than 1. These constraints facilitate good performance at short block-lengths because limiting both the density and the number of multiple edges in the protograph helps the resulting LDPC codes have good girth and avoid having too many short cycles. Explicit constraints are necessary because increasing the value of any element at any position of a protomatrix with a finite upper bound of \eqref{eq_first_theorem} or \eqref{eq_second_theorem} either results in an increase in the upper bound or the upper bound stays the same. 
\end{remark}

\begin{remark}
\label{remark_channel_used}
Our design method, which maximizes an upper bound on code minimum distance, does not depend upon the channel over which we deploy the codes. For designing codes for comparison according to the original design method that involves computing iterative decoding thresholds, we assume the binary-input additive white Gaussian noise channel (BI-AWGNC). The threshold values computed in this work are the result of at least 1000 iterations of the reciprocal channel approximation method to computing thresholds (see \cite{7045568}). 
\end{remark}

\begin{figure}[t]
\centerline{\includegraphics[width=3.4in]{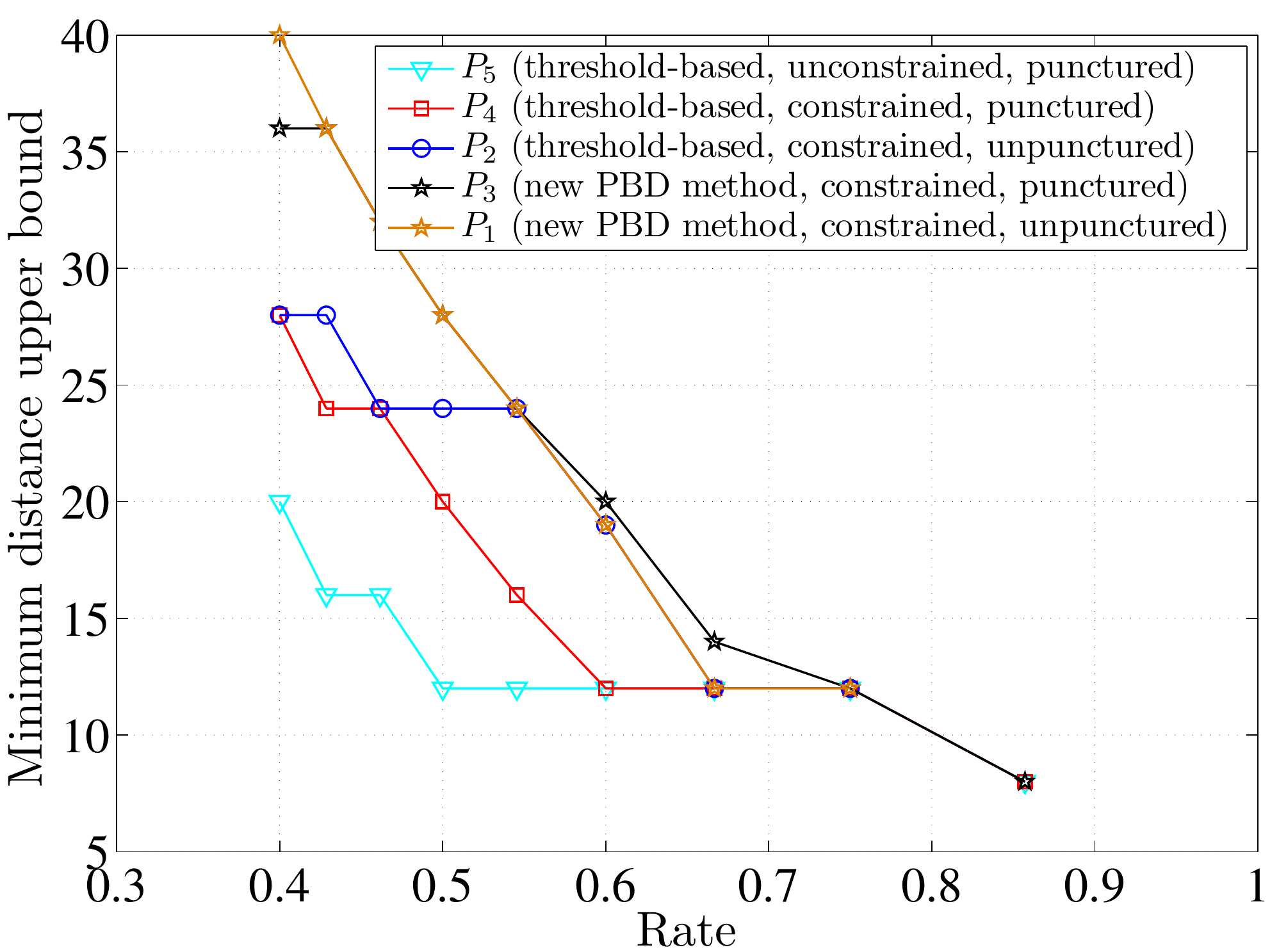}}
\caption{Comparison of minimum distance upper bounds.}
\label{fig_mind_comparison}
\end{figure}

\subsubsection{Unpunctured design} We design two unpunctured ensembles using the $\mathsf{HRC}$ matrix given in \eqref{eq_chosen_HRC} with the design constraints in Remark \ref{remark_design_constraints}. The design rates we consider decrease from $6/8$ to $6/15$. The first ensemble, $P_1$, is obtained using the new permanent bound design (PBD) method proposed in this paper. For comparison, the second ensemble, $P_2$, is designed by optimizing the iterative decoding thresholds (referred to as ``threshold-based'' in results) over BI-AWGNC.
The $\mathsf{IRC}$ parts of $P_1$ and $P_2$ that we obtained are:
\begin{align}
\small
\label{eq_IRC_parts_unpunctured}
P_{1,\mathsf{IRC}} = \begin{bmatrix}
1~1~1~0~0~0~0~0\\
0~0~0~1~1~1~0~0\\
1~0~0~0~0~0~1~1\\
0~1~1~1~0~0~0~0\\
1~0~0~0~1~1~0~0\\
1~0~1~0~0~0~0~1\\
0~1~0~1~0~0~1~0
\end{bmatrix},
P_{2,\mathsf{IRC}} = \begin{bmatrix}
1~1~1~0~0~0~0~0\\
0~0~0~1~1~1~0~0\\
1~0~0~0~0~0~1~1\\
0~0~1~0~1~0~1~0\\
1~0~0~1~1~0~0~0\\
1~0~1~0~0~1~0~0\\
1~0~1~0~0~0~1~0
\end{bmatrix}
\end{align}
\subsubsection{Punctured design} Similarly, we design two punctured ensembles using the $\mathsf{HRC}$ matrix in \eqref{eq_chosen_HRC} via the design constraints in Remark \ref{remark_design_constraints}. The first variable node is punctured\footnote{A punctured variable node improves the iterative decoding threshold (see \cite{5174517} and \cite{7045568}). Also, following the observations of \cite{7045568} we constrain every row of $P_{\mathsf{IRC}}$ to connect to the punctured variable node in our punctured designs.}, and the design rates decrease from $6/7$ to $6/15$. The resulting ensembles are called $P_3$ (PBD) and $P_4$ (threshold-based). 
Their $\mathsf{IRC}$ parts are
\begin{align}
\small
\label{eq_IRC_parts_punctured}
P_{3,\mathsf{IRC}}= \begin{bmatrix}
1~1~1~0~0~0~0~0\\
1~0~0~1~0~0~1~0\\
1~0~0~0~1~0~0~1\\
1~0~1~0~0~1~0~0\\
1~1~0~1~0~0~0~0\\
1~0~0~0~1~1~0~0\\
1~0~0~0~1~0~0~1\\
1~1~0~0~1~0~0~0
\end{bmatrix},P_{4,\mathsf{IRC}}= 
\begin{bmatrix}
1~0~1~0~1~0~0~0\\
1~0~1~0~0~0~1~0\\
1~1~0~0~1~0~0~0\\
1~0~0~1~0~0~1~0\\
1~0~0~0~1~1~0~0\\
1~0~0~0~0~0~1~1\\
1~0~0~1~1~0~0~0\\
1~1~0~0~0~0~1~0
\end{bmatrix}
\end{align}
\subsubsection{Unconstrained design via original PBRL design method} For further comparison, we design an ensemble, called $P_5$, using the same $\mathsf{HRC}$ matrix with its first variable node punctured according to the original PBRL design method. The design rates decrease from $6/7$ to $6/15$. For this ensemble we only have the following restriction in the $\mathsf{IRC}$ part: We do not permit any integer greater than 1.
The design yielded the following $\mathsf{IRC}$ part:
\begin{align}
\small
\label{eq_IRC_part_unconstrained}
P_{5,\mathsf{IRC}} =
\begin{bmatrix}
1~0~1~0~1~0~0~0\\
1~0~1~0~1~0~1~0\\
1~1~1~0~1~0~1~0\\
1~1~1~0~1~0~1~0\\
1~1~1~0~1~0~1~0\\
1~0~1~1~1~0~0~0\\
1~1~1~0~1~0~1~0\\
1~0~1~1~0~0~0~0
\end{bmatrix}
\end{align}

\begin{figure}[t]
\centerline{\includegraphics[width=3.4in]{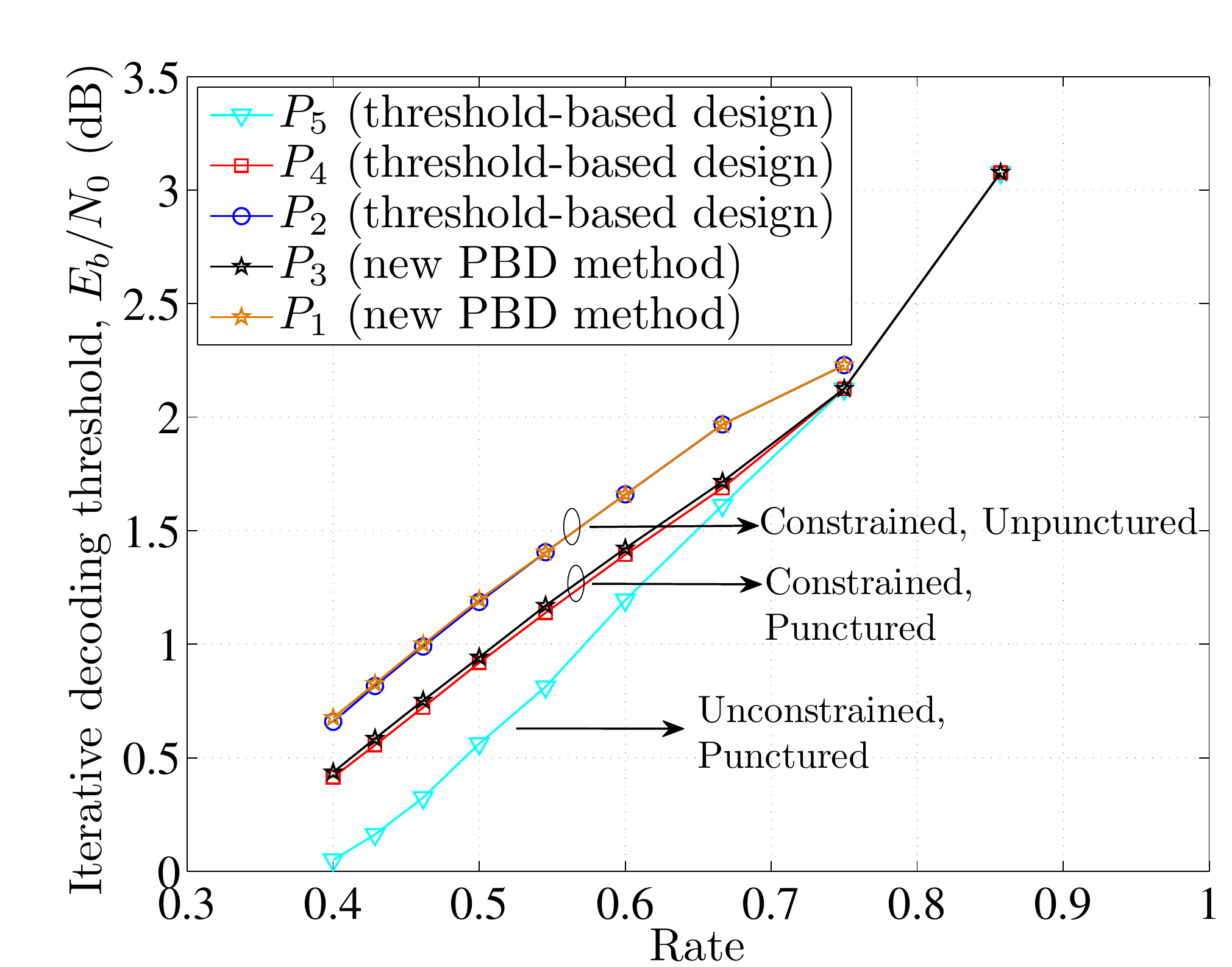}}
\caption{Comparison of iterative decoding thresholds over BI-AWGNC.}
\label{fig_threshold_comparison}
\end{figure}

We now compare the five ensembles $P_i, i \in [5]$ according to two metrics. Fig.~\ref{fig_mind_comparison} shows the upper bound on the minimum distance obtained at each rate for the five ensembles. $P_1$, the unpunctured, constrained ensemble obtained via the new PBD method, has the best upper bound at almost every rate. At the other end of the spectrum, $P_5$, the unconstrained, punctured ensemble designed to optimize the threshold at each rate, has the worst upper bound at every rate. 

The iterative decoding thresholds at each rate (over BI-AWGNC) for all five ensembles are shown in Fig.~\ref{fig_threshold_comparison}. As expected, ensemble $P_5$ has the best threshold at each rate. But surprisingly, both the unpunctured and punctured constrained ensembles obtained via the new PBD method, $P_1$ and $P_3$, have almost the same threshold at each rate as their counterpart ensembles, $P_2$ and $P_4$, which were obtained by optimizing the iterative decoding threshold at each rate. 
\section{Simulation Results}
\label{sec_results}
\begin{figure}[t]
\centerline{\includegraphics[width=3.5in]{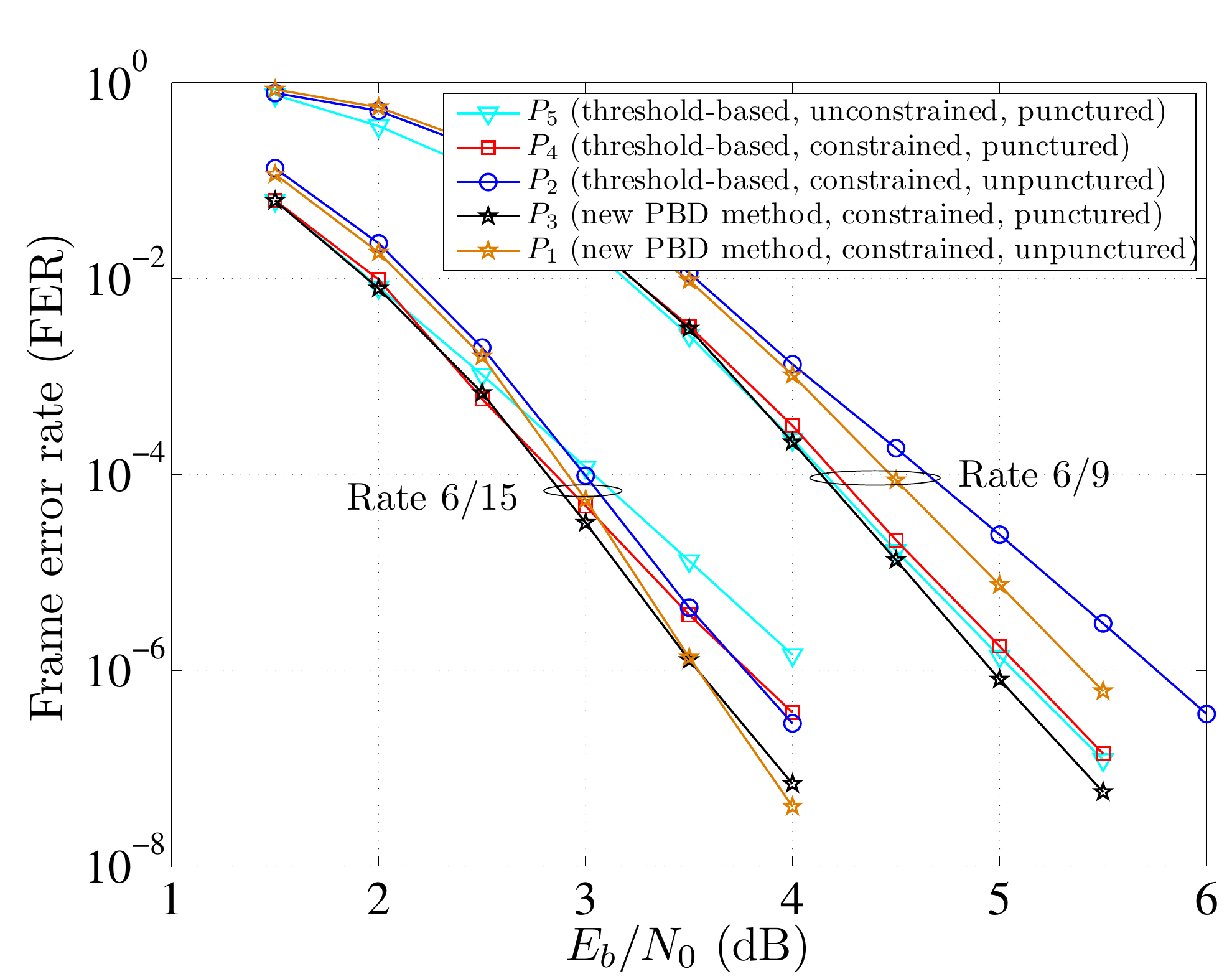}}
\caption{Comparison of FER performance at rates $6/15$ and $6/9$}
\label{fig_error_rate_comparison}
\end{figure}

\begin{figure}[t]
\centerline{\includegraphics[width=3.5in]{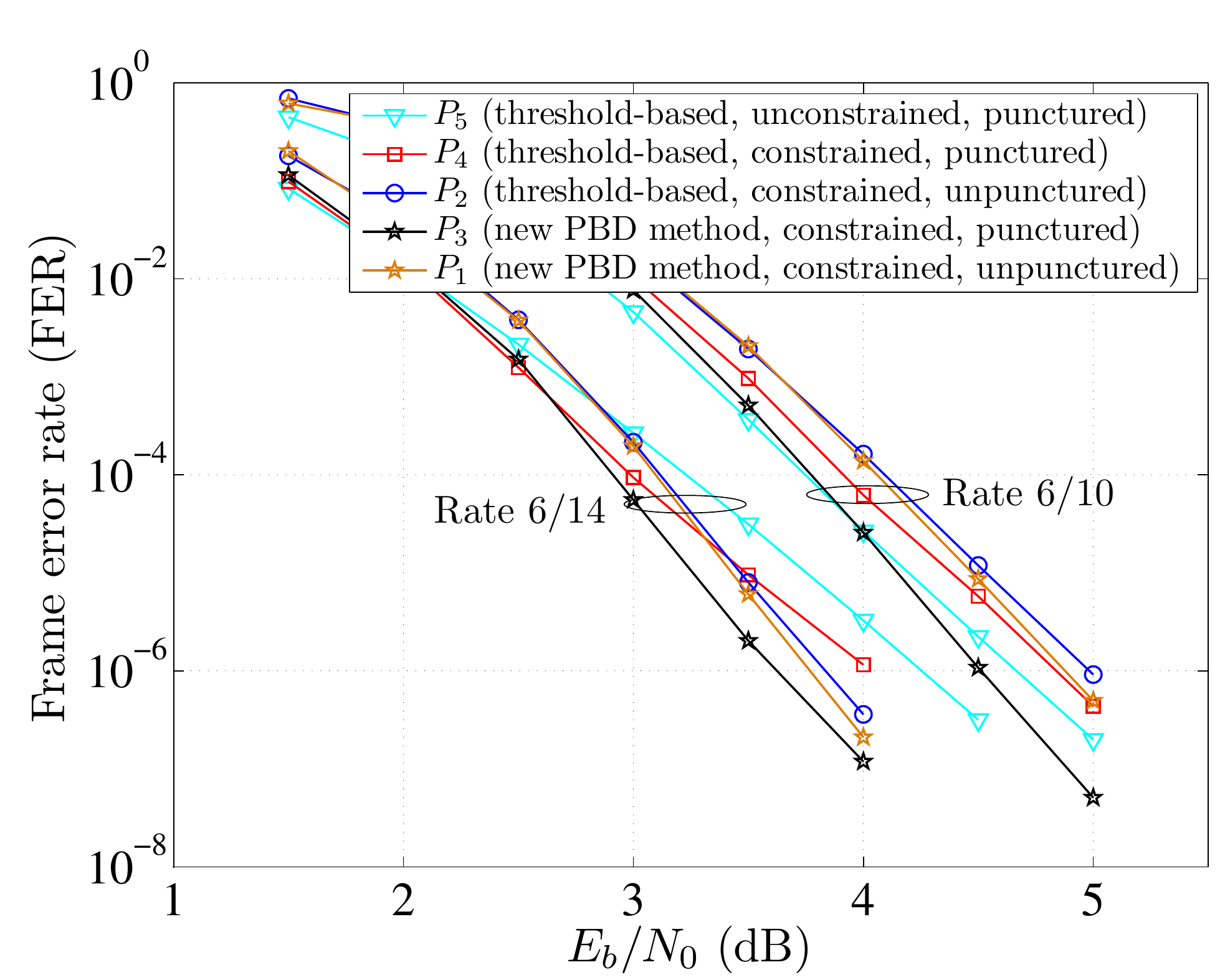}}
\caption{Comparison of FER performance at rates $6/14$ and $6/10$}
\label{fig_error_rate_comparison_1}
\end{figure}

This section presents simulation results of carefully designed RC code families from each of the five ensembles. Codes simulated in this section are all quasi-cyclic. Lifting was performed for the $n_c \times n_v$ protomatrix of the lowest rate $6/15$ using the circulant-PEG (C-PEG) algorithm combined with the ACE algorithm of Tian et al.\ \cite{1327837}. The lifting factor used is 33, which resulted in $k=198$ information bits for all code families. The resulting girth of all code families, at the lowest rate, is 6. Simulation results shown were obtained using a maximum of 100 iterations of full-precision, flooding, LLR-domain belief propagation over BI-AWGNC. At least 100 errors were collected for each frame error rate (FER) point in any simulated $\text{E}_{\text{b}}/\text{N}_{\text{0}}$ vs.\ FER graph.


The FER of all five codes are shown in Figs.~\ref{fig_error_rate_comparison} and \ref{fig_error_rate_comparison_1}. The QC-LDPC code family obtained from ensemble $P_3$ outperforms all the other codes at FERs $10^{-4}$, $10^{-5}$, and $10^{-6}$ and at all rates (even at rates $6/12$ and $6/13$, which are omitted as the results are similar to the ones shown here). This ensemble has the advantages of a good, if not the best, threshold due to the punctured variable node (Fig.~\ref{fig_threshold_comparison}) and a good upper bound on the minimum distance at all rates (Fig.~\ref{fig_mind_comparison}). The code family from the ensemble $P_1$, which has the best minimum distance upper bound at all rates, performs well at lower rates but not at higher rates. 
\begin{figure}[t]
\centerline{\includegraphics[width=3.1in]{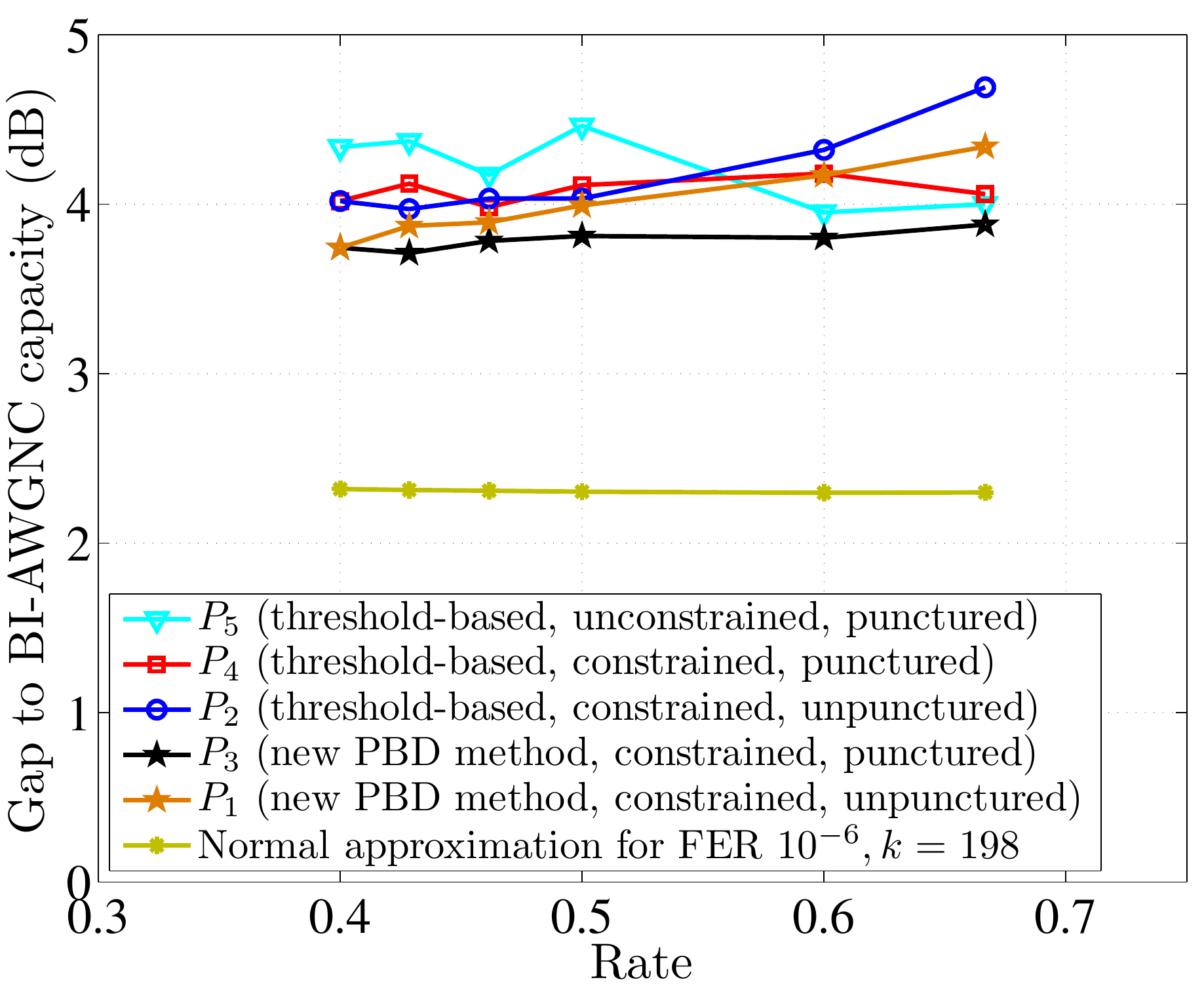}}
\caption{Comparison of gap to BI-AWGNC capacity at a frame error rate (FER) of $10^{-6}$ and information block size of $k=198$ for all the five codes and the refined normal approximation of \cite{5452208}}
\label{fig_norm_comparison}
\end{figure}

The gap to BI-AWGNC capacity at FER of $10^{-6}$ is shown in Fig.~\ref{fig_norm_comparison}. The code family of ensemble $P_3$ achieves the best performance at all design rates. The performance of this code is about $1.5$ dB away at all rates from the refined normal approximation of \cite{5452208}.
\section{Conclusion}
\label{sec_conclusion}
This paper proposed a new method to design PBRL QC-LDPC codes for short block-lengths. The metric used in the design is an upper bound on the minimum distance of any QC-LDPC code that can be obtained from a protomatrix. By maximizing this upper bound at each design rate of the rate-compatible family of codes, the paper obtained a significant improvement in the error floor region over PBRL codes designed according to the original method of optimizing the iterative decoding threshold. Furthermore, the paper identified a key reduction that is possible in the complexity of the newly proposed design procedure. 
\bibliographystyle{IEEEtran}
\bibliography{IEEEabrv,mybib}
\appendix[Reduction in Complexity of Design Procedure for PBRL Protomatrices With Punctured Nodes]
\label{sec_appendix}
Theorem \ref{theorem_main_result} provides a considerable reduction in the complexity of the new design procedure to obtain PBRL ensembles with no punctured variable nodes. This appendix explores the case when a protomatrix has punctured variable nodes. 

Let a PBRL protomatrix $P$ of size $n_c \times n_v$ with $n_p$ punctured variable nodes have a positive design rate less than 1, i.e.\ $n_v > n_c$ and $n_t = n_v - n_p > n_v - n_c$. Let the $\mathsf{HRC}$ part be of size $n_{c_H} \times n_{v_H}$. Let the set of punctured variable nodes be denoted $\mathcal{P}$. Assume that the upper bound in $\eqref{eq_second_theorem}$ for $P$ is a positive integer. 

Now, if $S \subseteq \left[n_v\right]$, $|S| = n_c + 1$, $S \cap \mathcal{P} = \phi$, and $S$ does not contain all the columns of $P_{\mathsf{IR}}$, then the arguments in Theorem \ref{theorem_main_result} for ignoring such a set $S$ while computing the upper bound for the protomatrix still hold. Similarly, if $S$ includes $P_{\mathsf{IR}}$ and any subset of columns from the first $n_{v_H}$ columns of $P$, the computational complexity arguments for computing the sum of at most $n_c + 1$ permanents for such a set of columns hold the same way as observed in Theorem \ref{theorem_main_result}. 

Now consider the case when $S \subseteq \left[n_v\right]$, $|S| = n_c + 1$, $S$ does not contain all the columns of $P_{\mathsf{IR}}$, and $S \cap \mathcal{P} \ne \phi$. Let the number of punctured columns in $S$ be $n_{S_p}$. The sum
\begin{align}
\label{eq_appendix_equation_1}
\sum_{i \in S \setminus \mathcal{P}} \mathsf{perm}\left(P_{S \setminus i}\right)
\end{align}
has $n_c + 1 - n_{S_p}$ permanents that need to be computed, all of size at most $n_c \times n_c$. Every $n_c \times n_c$ sub-matrix whose permanent is computed for this sum contains all the $n_{S_p}$ columns that are punctured. The strategy of replacing $n_v - n_{v_H}$ columns of a sub-matrix whose permanent is non-zero in \eqref{eq_appendix_equation_1} by all the columns in $P_{\mathsf{IR}}$ does not necessarily yield a new summation that is smaller than the summation computed for the columns in $S$ according to \eqref{eq_appendix_equation_1}. We give an example to illustrate this. 

Let us consider the following PBRL protomatrix whose first column is punctured:
\begin{align}
\label{eq_appendix_equation_2}
\begin{bmatrix}
1&1&2&1&2&1&0\\
0&2&1&2&1&2&0\\
1&1&0&0&0&0&1
\end{bmatrix}
\end{align}
Let $S=\{1,2,3,4\}$, the first four columns of the protomatrix. The sum in \eqref{eq_appendix_equation_1} for this set has three terms in it and is equal to 17. The permanent of the $3 \times 3$ sub-matrix comprised of columns $1, 3, 4$ is equal to 5. If we follow the same replacement strategy as in Theorem \ref{theorem_main_result}, we need to replace column 1 by the only incremental redundancy variable node, i.e.\ column 7. The new set of columns $\{2, 3, 4, 7\}$ has no column that is punctured and the summation in \eqref{eq_appendix_equation_1}, which now has four terms, is now equal to 19. 

Therefore, we observe that when a PBRL protomatrix has $n_p$ punctured variable nodes and a finite upper bound in \eqref{eq_second_theorem}, the complexity of computing the bound can be reduced the following way: 
\begin{enumerate}
\item Consider all sets of $n_c + 1$ columns that contain $P_{\mathsf{IR}}$. For such a set we need to compute at most $n_c + 1$ permanents, each of size at most $\left(n_{c_H} + 1\right) \times \left(n_{c_H} + 1\right)$. There are $\binom{n_{v_H}}{n_{c_H} + 1}$ such sets.
\item Consider all sets of $n_c + 1$ columns that contain $i \ge 1$ punctured variable nodes, and not all columns or no column of $P_{\mathsf{IR}}$. Compute the summation of the $n_c + 1 - i$ permanents, each of size at most $n_c \times n_c$ (if there are columns of $P_{\mathsf{IR}}$ then the complexity of computing the permanent would decrease). There are
\begin{align}
\label{eq_appendix_equation_3}
\sum_{i=1}^{n_p} \sum_{j=0}^{n_v - n_{v_H} - 1} \binom{n_p}{i}\binom{n_{v_H} - n_p}{n_c + 1 - i - j} \binom{n_v - n_{v_H}}{j}
\end{align}
such sets that need to be considered, where the binomial coefficient $\binom{n}{k}$ is assumed to be 0 if $k \notin \{0,1,\dots,n\}$. 
\end{enumerate}
\end{document}